\documentclass[12pt]{article}

\usepackage{fullpage}
\usepackage{graphicx}
\usepackage{amsmath}
\usepackage{url}
\usepackage{parskip}
\usepackage{hyperref}
\usepackage[margin=15pt,font=small,labelfont=bf,labelsep=endash]{caption}
\hypersetup{
    colorlinks=true,
    linkcolor=black,
    citecolor=black,
    filecolor=black,
    urlcolor=black,
}

\newcommand{\smat}{\left( \begin{matrix}} 
\newcommand{\emat}{\end{matrix} \right)} 

\newcommand{\bea}{\begin{eqnarray*}} 
\newcommand{\eea}{\end{eqnarray*}} 

\newcommand{\beq}{\begin{equation}} 
\newcommand{\be}[1]{\begin{equation}\label{#1}} 
\newcommand{\ee}{\end{equation}} 

\newcommand{\bskip}{\bigskip\noindent} 

\renewcommand{\Im}{\operatorname{Im}}
\renewcommand{\Re}{\operatorname{Re}}

\newcommand{\E}{\mathbf{E}}
\renewcommand{\H}{\mathbf{H}}
\renewcommand{\k}{\mathbf{k}}

\renewcommand{\S}{\mathbf{S}}
\renewcommand{\r}{\mathbf{r}}
\newcommand{\x}{\hat{\mathbf{x}}}
\newcommand{\y}{\hat{\mathbf{y}}}
\newcommand{\z}{\hat{\mathbf{z}}}

\begin{document}
\title{Multilayer optical calculations}
\date{\small{\today}}
\author{Steven J. Byrnes \\ \small{Current affiliation: Charles Stark Draper Laboratory, Cambridge, Massachusetts, USA} \\ \small{Contact: steven.byrnes@gmail.com}}
\maketitle
\abstract{When light hits a multilayer planar stack, it is reflected, refracted, and absorbed in a way that can be derived from the Fresnel equations. The analysis is treated in many textbooks, and implemented in many software programs, but certain aspects of it are difficult to find explicitly and consistently worked out in the literature. Here, we derive the formulas underlying the transfer-matrix method of calculating the optical properties of these stacks, including oblique-angle incidence, absorption-vs-position profiles, and ellipsometry parameters. We discuss and explain some strange consequences of the formulas in the situation where the incident and/or final (semi-infinite) medium are absorptive, such as calculating $T>1$ in the absence of gain. We also discuss some implementation details like complex-plane branch cuts. Finally, we derive modified formulas for including one or more ``incoherent'' layers, i.e. very thick layers in which interference can be neglected. This document was written in conjunction with the ``tmm'' Python software package, which implements these calculations.}

\newpage
\tableofcontents

\section{Introduction}

I originally wrote these notes to record and explain the calculations implemented by the ``tmm'' (short for ``transfer matrix method'') Python software package: See \url{https://pypi.python.org/pypi/tmm}.

The derivations (at least through Sec.~\ref{Sec:MultilayerThinFilms}) can be found, in whole or part, in quite a few textbooks and references. I found Bo Sernelius's lecture notes\footnote{\url{http://people.ifm.liu.se/boser/elma/}--especially lecture 13} an especially useful starting point.

Apart from my tmm program, there are many other programs that calculate some or all of the same formulas.\footnote{I have a list at: \url{http://sjbyrnes.com/multilayer-film-optics-programs/}} I have done a few consistency checks between my program and others. They tend to agree perfectly except in the tricky (and somewhat unusual) case of calculating reflected power or transmitted power when the semi-infinite incoming and/or outgoing medium has a complex index of refraction.

I assume non-magnetic ($\mu=\mu_0$) and isotropic (as opposed to birefringent) materials throughout the document.

\section{Wave propagation}

We assume our structure is a stack of one or more smooth planar layers, such as a flat piece of glass with an antireflective coating on top. The interfaces between layers are all normal to $\z$, and everything is uniform in the $x$ and $y$ directions. We assume the wavevector of the light is in the $x$--$z$ plane. (Or just along $\z$ if it's normal-incidence). The ``forward'' direction (direction that normally-incident incoming light is traveling) is $+\z$.

All sinusoidally-oscillating quantities are given as complex numbers; to get the actual value at any particular time, multiply by $e^{-i \omega t}$ and take the real part.

The electric field at any given point is a superposition of the forward-moving and backwards-moving electromagnetic waves:
\beq\E(\r) = \E_f^0 e^{i\k_f\cdot\r} + \E_b^0 e^{i\k_b\cdot\r}\ee
Here, $\k_f$ and $\k_b$ are the [angular] wavevectors for forward- and backwards-moving waves; $\E_f^0$ and $\E_b^0$ are some constant vectors; and $\E(\r)$ is the complex electric field at any given point $\r$ within a certain layer. The $y$-components of the $\k_f$ and $\k_b$ are zero because, like I said above, the wavevector is assumed to be in the $x$--$z$ plane. The $x$-components of $\k_f$ and $\k_b$ are always real, because we are assuming it's a plane wave, so the light intensity is uniform along the $x$ and $y$ directions. However, the $z$ component might be complex, representing a wave that is attenuating as it travels through the stack, due to absorption.

The wavevectors is related to the [complex] index of refraction $n$ by:
\beq
\k_f = \frac{2\pi n}{\lambda_{vac}}(\z \cos \theta + \x \sin \theta) \quad , \quad \k_b = \frac{2\pi n}{\lambda_{vac}}(-\z \cos \theta + \x \sin \theta)
\ee
where $\theta$ is the angle from the normal, and $\lambda_{vac}$ is the vacuum wavelength. That means $n \sin \theta$ is always a real number, but $n \cos \theta$ might not be. This is consistent with Snell's law:
\beq n_i \sin \theta_i = n_j \sin \theta_j\ee
i.e., $n\sin \theta$ should be the same real number in every layer. Snell's law is the same as saying that the component of $\k$ in the $x-y$ plane is the same in each layer.

\subsection{What is complex refractive index?}

When the refractive index is complex, the imaginary part is sometimes called ``extinction coefficient''. The larger it is, the more quickly light gets absorbed as it tries to travel through the material. Negative extinction coefficient corresponds to stimulated emission. Extinction coefficient, like refractive index, is a unitless number. It should \emph{NOT} be confused with ``molar extinction coefficient'' or ``mass extinction coefficient'' in chemistry, which are not unitless.\footnote{However, if you know one you can figure out the other. See \url{https://en.wikipedia.org/w/index.php?title=Mathematical_descriptions_of_opacity&oldid=695422502}.} For real-world materials, the extinction coefficient, like the refractive index, is different at different frequencies.

Again, with the conventions used here, $\Im n>0$ means absorption and $\Im n<0$ means stimulated emission.

\subsection{Explicit \texorpdfstring{$\E$, $\H$, and $\k$}{E, H, and k} for s-polarization and p-polarization}

As usual, $s$-polarization is where the $\E$-field points in the $y$-direction, and $p$-polarization is where the $\H$-field points in the $y$-direction. There is no difference between $s$ and $p$-polarization for normal-incident light. However, the way our sign conventions work, the reflection amplitude for a normal-incidence wave is given with the opposite sign depending on whether you call it ``s'' or ``p``! For $s$-polarization, our sign convention is based on the direction of the $E$-field---e.g., if two waves have $E$ pointing parallel, then their amplitudes have the same sign. Whereas for $p$-polarization, our sign convention is based on the direction of the $H$-field. Half of textbooks use the opposite sign convention for $p$ from this one, so don't be surprised to see discrepancies between different sources. See Appendix~\ref{appendixrp} for further discussion.

Maxwell's equations imply that $\H = \frac{1}{\mu_0\omega}\k \times \E$ for a plane wave. [This works even if $\k$ and/or $\E$ is a complex vector.] Therefore the explicit $\E,\H,\k$ are:

\begin{center} {\bf \underline{s-polarization}} \end{center}
\vspace{-5mm}
\begin{align}
\k_f =\frac{2\pi n}{\lambda_{vac}}\left( \cos \theta \z + \sin \theta \x\right) \quad , &\quad \k_b = \frac{2\pi n}{\lambda_{vac}}\left(- \cos \theta \z + \sin \theta \x\right) \nonumber \\
\E_f = E_f \y \quad , &\quad \E_b = E_b \y \nonumber \\
\H_f \propto  n E_f \left( - \cos \theta \x +  \sin \theta \z\right)\quad , &\quad \H_b \propto n E_b\left( \cos \theta \x + \sin \theta \z\right) \label{Eq:SWaves}
\end{align}

and \begin{samepage}
\begin{center} {\bf \underline{p-polarization}} \end{center}
\vspace{-5mm}
\begin{align}
\k_f =\frac{2\pi n}{\lambda_{vac}}\left( \cos \theta \z + \sin \theta \x\right) \quad , &\quad \k_b = \frac{2\pi n}{\lambda_{vac}}\left(- \cos \theta \z + \sin \theta \x\right) \nonumber \\
\E_f = E_f(-\sin \theta \z +  \cos \theta \x) \quad, &\quad \E_b = E_b(- \sin \theta \z -  \cos \theta \x) \nonumber \\
\H_f \propto n E_f \y \quad , &\quad \H_b \propto n E_b \y \label{Eq:PWaves}
\end{align} \end{samepage}

\section{Single-interface reflection and transmission amplitudes}

If you have the interface between two layers 1 and 2, and shine light from 1, then there are three relevant wave amplitudes: The incident amplitude ($E_f$ on the layer 1 side), the reflected amplitude ($E_b$ on the layer 1 side), and the transmitted amplitude ($E_f$ on the layer 2 side). The reflection coefficient $r$ is the ratio of reflected amplitude to incident amplitude, and the transmission coefficient $t$ is the ratio of transmitted amplitude to incident amplitude.

To derive the equations for $r$ and $t$ (``The Fresnel Equations''), we start with Eqs.~(\ref{Eq:SWaves}-\ref{Eq:PWaves}) on both sides of the interface, with $E_f=E_0, E_b=rE_0$ on the starting side and $E_f=tE_0, E_b=0$ on the destination side. Then we plug in the boundary conditions for $E$ and $H$: $H_x,H_y,H_z,E_x,E_y,$ and $n^2 E_z$ are all continuous across the boundary. Two clarifications: (1) These boundary conditions refer to components of the \emph{total} fields $\E=\E_f + \E_b$ and $\H=\H_f+\H_b$; (2) The criterion is actually that $B_z$ and $D_z$ are continuous across the boundary, but I just rephrased it in terms of $H_z$ and $E_z$ using $\mu=\mu_0, \epsilon=n^2\epsilon_0$.

Readers may check the algebra themselves, or refer to any optics textbook; traveling from medium 1 into medium 2:
\begin{align}
r_s = \frac{n_1 \cos \theta_1 - n_2 \cos \theta_2}{n_1 \cos \theta_1 + n_2 \cos \theta_2} \quad &, \quad r_p = \frac{n_2 \cos \theta_1 - n_1 \cos \theta_2}{n_2 \cos \theta_1 + n_1 \cos \theta_2} \nonumber \\
t_s = \frac{2n_1 \cos \theta_1}{n_1 \cos \theta_1 + n_2 \cos \theta_2} \quad &, \quad t_p = \frac{2n_1 \cos \theta_1}{n_2 \cos \theta_1 + n_1 \cos \theta_2} \label{Eq:fresnel}
\end{align}

\subsection{Waves coming from both sides}
A slight extension is to have waves incoming from both sides of the interface at once. Say they have amplitudes (at the interface) of $E_{f1},E_{b1},E_{f2},E_{b2}$ (forwards in medium 1, backwards in 1, forwards in 2, backwards in 2). These are related by
\be{Eq:4wave} E_{b1} = E_{f1}r_{12} + E_{b2}t_{21} \quad , \quad E_{f2} = E_{f1}t_{12} + E_{b2}r_{21}\ee
where $t_{ab},r_{ab}$ are transmission and reflection going from layer $a$ into $b$. Intuitively, each of the outgoing waves is a superposition of the reflected amplitude of one of the incoming waves, plus the transmitted amplitude of the other incoming wave. More formally, we could derive these from scratch using the electromagnetic boundary conditions, but it's easier to note that we already know two situations that satisfy the boundary conditions: $(E_{f1},E_{b1},E_{f2},E_{b2}) = (1,r_{12},t_{12},0)$ and $(0,t_{21},r_{21},1)$.  Both of these satisfy (\ref{Eq:4wave}), and therefore so does any linear combination / superposition of those two situations---which includes all possible combinations of incoming waves.

\section{Multilayer thin films \label{Sec:MultilayerThinFilms}}
\subsection{Complex amplitudes for reflection and transmission}

\begin{figure}[htb]
\centering
\includegraphics[width=0.5\textwidth]{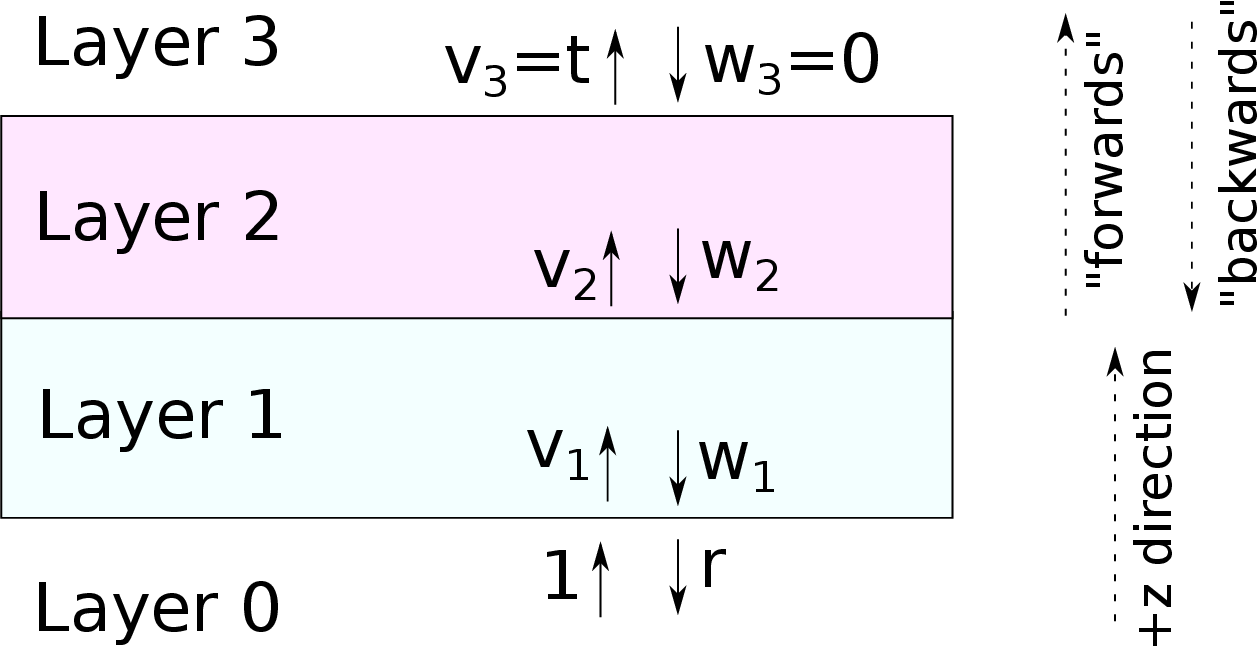}
\caption{Sample stack with $N=4$ (two finite layers between two semi-infinite layers). The labels next to the small arrows indicate wave amplitudes.\label{vwdefn}}
\end{figure}

Now we have $N$ materials, numbered $0,1,\ldots,N-1$, where the first (``0'') and last (``$N-1$'') layer are semi-infinite. Light with amplitude 1 is in layer 0, heading towards layer 1 (Fig.~\ref{vwdefn}).

At the interface between the $(n-1)$st and $n$th material, let $v_n$ be the amplitude of the wave on the $n$th side heading forwards (away from the boundary), and let $w_n$ be the amplitude on the $n$th side heading backwards (towards the boundary). (Fig.~\ref{vwdefn}.) (($v_0,w_0$) are undefined, while $v_{N-1}=t$ and $w_{N-1}=0$.) We define
\beq \delta_n \equiv (\text{thickness of layer }n)(k_z\text{ for the forward-traveling wave in layer }n)\ee
i.e., $\delta_n$ characterizes the phase [and when $k_z$ is complex, also the absorption] that comes from passing through layer $n$. Now from Eq.~(\ref{Eq:4wave}) we get:
\begin{eqnarray}
v_{n+1} &=& (v_n e^{i \delta_n})t_{n,n+1} + w_{n+1}r_{n+1,n} \nonumber \\
w_n e^{-i\delta_n} &=& w_{n+1} t_{n+1,n} + (v_n e^{i \delta_n})r_{n,n+1} 
\end{eqnarray}
where $r_{a,b}$ and $t_{a,b}$ are reflection and transmission for light heading from layer $a$ into layer $b$. Using the identities $r_{a,b} = -r_{b,a}$ and $t_{a,b}t_{b,a} - r_{a,b}r_{b,a} = 1$ (which follow from Eqs.~(\ref{Eq:fresnel})), we can transform these into:
\beq \smat v_n \\ w_n \emat = M_n \smat v_{n+1}\\w_{n+1}\emat\ee
for $n=1,\ldots,N-2$, where
\beq M_n \equiv \smat e^{- i \delta_n} & 0 \\ 0 & e^{i \delta_n} \emat \smat 1 & r_{n,n+1} \\ r_{n,n+1} & 1 \emat \frac{1}{t_{n,n+1}}\ee
Now we want the matrix relating the waves entering the structure to the waves exiting, i.e.:
\beq \smat 1 \\ r \emat = \tilde{M} \smat t \\ 0 \emat.\ee
$\tilde{M}$ is given by:
\beq \tilde{M} = \frac{1}{t_{0,1}} \smat 1 & r_{0,1} \\ r_{0,1} & 1 \emat M_1 M_2 \cdots M_{N-2}\ee
Combining these two equations allows $r$ and $t$ to be written in terms of the four entries of the matrix $\tilde{M}$:
\beq \smat 1\\r \emat = \smat \tilde{M}_{00} & \tilde{M}_{01} \\ \tilde{M}_{10} & \tilde{M}_{11} \emat \smat t \\ 0 \emat\ee
\beq t = 1/\tilde{M}_{00}, \quad r = \tilde{M}_{10}/\tilde{M}_{00} \ee
So now we know how to calculate $r$ and $t$ for an arbitrary multi-layer thin film. (And incidentally, it is straightforward from here to also calculate $v_n$ and $w_n$ for every $n$.)

\subsection{Ellipsometric parameters}

If we know $r_s$ and $r_t$, we can also calculate the two parameters measured in ellipsometry:
\beq \psi \equiv \tan^{-1}(|r_p/r_s|) \quad , \quad \Delta \equiv \operatorname{phase}(-r_p/r_s)\ee
where ``phase'' means complex phase angle.

However, I found that different textbooks have different definitions. So you may need to flip the signs, add or subtract $\pi/2$, etc.

\subsection{Calculating Poynting vector}

\begin{figure}[htb]
\centering
\includegraphics[width=0.5\textwidth]{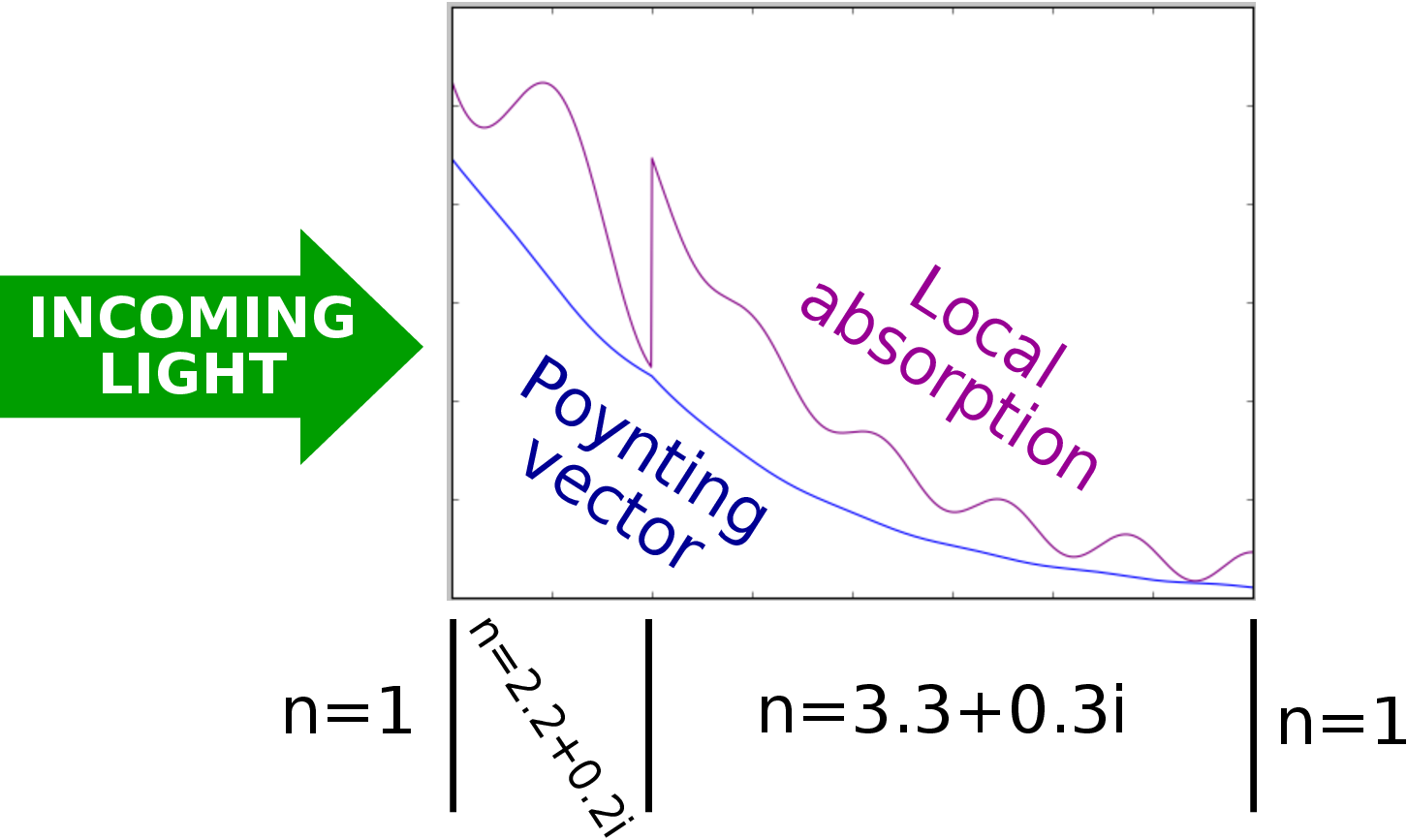}
\caption{Sample calculation of local absorption and Poynting vector in a two-layer structure with air on both sides (refractive indices written below the graph).\label{spatiallyresolvedexample}}
\end{figure}

The next few sections relate to power flows and power absorption: The goal is to be able to generate graphs like Fig.~\ref{spatiallyresolvedexample}. The relevant equations are somewhat hard to find (without typos) in the literature, but I verified them by various consistency checks, such as continuity across interfaces when appropriate, agreement with $R$ and $T$ in simple cases, etc.

I will start by deriving the expression for the normal component of the Poynting vector $\S$, i.e. $\S\cdot\z$. This dot-product represents the net power flowing forward through the structure at a given point. We compute it as a unitless fraction of the total incoming power. Start with s-polarization, using Eq.~(\ref{Eq:SWaves}):

\begin{eqnarray}
\E &=& E_f \y + E_b \y \nonumber \\
\H &\propto& n E_f \left( - \cos \theta \x +  \sin \theta \z\right) + n E_b\left( \cos \theta \x + \sin \theta \z\right) \nonumber \\
\S \cdot \z &=& \frac12 \Re[ \z \cdot (\E^* \times \H)] \propto \Re[(E_f^*+E_b^*) (E_f - E_b)n \cos \theta]
\end{eqnarray}
I am really only interested in power flow as a fraction of incoming power. The incoming power is what you would get with $E_f=1,E_b=0$. So here is the final result:
\be{Eq:PoyntingS} \text{s-polarization:} \qquad \S \cdot \z = \frac{\Re\left[ (n) (\cos \theta) (E_f^* + E_b^*) (E_f - E_b)\right]}{\Re\left[ n_0 \cos \theta_0\right] }\ee

Next, p-polarization, from Eq.~(\ref{Eq:PWaves}):
\begin{eqnarray}
\E &=& E_f(-\sin \theta \z +  \cos \theta \x) + E_b(- \sin \theta \z -  \cos \theta \x) \nonumber \\
\H &\propto& n E_f \y + n E_b \y \nonumber \\
\S \cdot \z &=& \frac12\Re[ \z \cdot (\E^* \times \H)] \propto \Re[(\cos \theta)^*(E_f^*-E_b^*) (E_f + E_b)n ]\end{eqnarray}
Again we normalize to incident power:
\be{Eq:PoyntingP} \text{p-polarization:} \qquad \S \cdot \z = \frac{\Re\left[ (n) (\cos \theta^*) (E_f + E_b) (E_f^* - E_b^*)\right]}{\Re \left[ n_0 \cos \theta_0^* \right]}\ee

(If I omitted parentheses somewhere, it's because $\cos (\theta^*) = (\cos \theta)^*$.)

\subsection{\texorpdfstring{$T$}{T} (transmitted power) and \texorpdfstring{$R$}{R} (reflected power)}

To get the formula for $T$, the fraction of power transmitted, we simply take Eqs.~(\ref{Eq:PoyntingS}),(\ref{Eq:PoyntingP}) and apply it to the final medium by plugging in $E_b=0$ (no light is flowing backwards in the final layer):
\beq \text{s-polarization:} \qquad T = |t|^2\frac{ \Re\left[ n \cos \theta\right]}{\Re \left[ n_0 \cos \theta_0 \right] }\ee
\beq \text{p-polarization:} \qquad T = |t|^2\frac{ \Re\left[ n \cos \theta^* \right]}{\Re \left[ n_0 \cos \theta_0^* \right] }\ee
where $T$ is the fraction of power transmitted and $t=E_f/E_0$ is the transmission amplitude.\footnote{In some references, the complex conjugation for p-polarization is omitted, but I'm very confident it's correct. Usually the incident and final media are non-absorbing, e.g. air, so $\cos \theta$ is real and it doesn't matter whether you conjugate $\theta$ or not.}

The formula for $R$ is just what you expect:
\beq R=|r|^2\ee

\subsubsection{Counter-intuitive results when the initial medium is absorptive}

When the initial medium is absorbing, these formulas give very strange results. Most strikingly, you can get $T>1$ in the absence of stimulated emission, and you can get $R+T > 1$ or $R+T < 1$ for an interface between two semi-infinite media. The issue more specifically is that the power entering the first layer of the stack (called \verb=power_entering= in the tmm software) is \emph{not} necessarily equal to $1-R$, as one would expect (energy 1 moving forwards, minus energy $R$ moving backwards). I explain and discuss this in Appendix~\ref{appendixR}.

Quick summary of Appendix~\ref{appendixR}: The difference between \verb=power_entering= and $1-R$---which can only happen when the starting medium is absorptive---is related to an excess or deficit of absorption just before the first interface, arising from interference between the incoming and reflected waves. In other words, $R$ and $T$ are normalized to incoming power far from the interface, extrapolated to the interface assuming exponential decay; but due to the interference, that extrapolation is inaccurate. So the actual power at the interface may be higher or lower than the power-normalization factor.

\subsection{Absorbed energy density}
Next, absorbed energy density at a given depth. In principle this has units of [power]/[volume], but we can express it as a multiple of incoming light power density on the material, which has units [power]/[area], so that absorbed energy density has units of 1/[length]. This is the negative derivative (with respect to distance) of the $\S \cdot \z $ expressions above. Differentiating is straightforward, using $E_f(z) \propto e^{i k_z z}$ and $E_b(z) \propto e^{-i k_z z}$. (Reminder: $k_z=2\pi n\cos\theta/\lambda_{vac}$.) The result is:
\begin{eqnarray}
\text{s-polarization:} \qquad a(z) &=& \frac{|E_f+E_b|^2 \Im\left[ n \cos(\theta) k_z\right] }{\Re\left[ n_0 \cos \theta_0 \right] } \nonumber \\
\text{p-polarization:} \qquad a(z) &=& \frac{\Im\left[ n \cos(\theta^*)\left(k_z|E_f-E_b|^2-k_z^*|E_f+E_b|^2\right)\right]}{\Re \left[ n_0 \cos \theta_0^* \right] }
\end{eqnarray}

Within a given layer, absorption is an analytical function:
$$a(z) = A_1e^{2 z \Im(k_z)} + A_2e^{-2 z \Im(k_z)} + A_3e^{2iz\Re(k_z)} + A_3^* e^{-2iz \Re(k_z)}$$
where:
\bea
\text{s-polarization}: A_1 &=& \frac{\Im\left[ n \cos(\theta) k_z\right] }{\Re\left[ n_0 \cos \theta_0 \right] }|w|^2\\
A_2 &=& \frac{\Im\left[ n \cos(\theta) k_z\right] }{\Re \left[ n_0 \cos \theta_0 \right] }|v|^2\\
A_3 &=& \frac{\Im\left[ n \cos(\theta) k_z\right] }{\Re \left[ n_0 \cos \theta_0 \right] }vw^*\\
\text{p-polarization}: A_1 &=& \frac{2\Im\left[ k_z\right] \Re\left[ n \cos(\theta^*)\right] }{\Re \left[ n_0 \cos \theta_0^* \right] } |w|^2\\
A_2 &=& \frac{2\Im\left[ k_z \right] \Re \left[ n \cos(\theta^*)\right] }{\Re \left[ n_0 \cos \theta_0^* \right] } |v|^2\\
A_3 &=& \frac{-2\Re\left[ k_z \right] \Im \left[ n \cos(\theta^*)\right] }{\Re \left[ n_0 \cos \theta_0^* \right] } vw^*
\eea
where $v = E_f(0)$ and $w=E_b(0)$. (For the purpose of this section, $z=0$ is the start of the layer in question. $n_0$ and $\theta_0$ refer as usual to the incident semi-infinite medium; remember, we are calculating absorption per unit incident power.)

\section{Branch cuts \label{branchcuts}}

Snell's law gives $\theta_i = \arcsin(n_0 \sin(\theta_0)/n_i)$. However, the arcsine function is ambiguous--it has branch cuts in the complex plane. How do we get the right $\theta$?

There are actually only two non-equivalent choices. If $\theta$ is one solution, then $\pi-\theta$ is the other. You may recognize that we are choosing which of the two waves in medium $i$ is called ``forward-traveling'' and which one is called ``backward-traveling''. How do we make the right choice?

\underline{Good news}: In the intermediate, finite-thickness layers, the choice actually doesn't matter. We solve for \emph{both} the forward- and backward-traveling waves, so it doesn't matter which wave has which name. The two choices of $\theta_i$ will switch $v_i$ with $w_i$, but won't affect observable quantities like reflectance, absorption, etc.

\underline{Bad news}: The choice of $\theta$ versus $\pi-\theta$ \emph{does} matter very much in the starting semi-infinite layer (where the ``forward-traveling'' wave amplitude is set to 1), and in the final semi-infinite layer (where the ``backwards-traveling'' wave amplitude is set to 0). In these layers, we need to choose $\theta$ correctly.

\underline{More bad news}: If you do the naive thing, $\theta_i=\arcsin(n_0 \sin(\theta_0)/n_i)$, you do not always wind up the right $\theta_i$. It depends on how arcsine is defined in your programming language of choice (branch cuts are inherently arbitrary). For example, Python/SciPy wants to choose the wrong $\theta$ for the final layer during total internal reflection. Therefore I recommend you always \emph{check} whether $\theta$ or $\pi-\theta$ is the right choice---see Appendix~\ref{appendixbranchcuts} for the specific criteria.

\section{Thick ``incoherent'' films}
\subsection{Introduction}
That finishes the discussion of thin-film interference. Next, thick films. Here we are interested in hybrid structures containing both thick and thin layers (or even just thick layers). Light loses its coherence when traveling through the thick layers---i.e., the Fabry--P\'{e}rot fringes are so close together that they cannot be resolved by the experimental measurement, due to factors such as random thickness variations, propagation angle variations, and/or wavelength variations. Instead of seeing the fringes, you just see the average.

I reiterate that an incoherent analysis is \emph{never strictly necessary}. If incoherence comes from having a variety of wavelengths / angles / thicknesses, the obvious way to proceed is to do many coherent simulations across a variety of wavelengths / angles / thicknesses and then average the results. However, the incoherent analysis is a convenient shortcut when appropriate.

In the tmm software approach, there are two types of layers: Coherent layers (treated as in the sections above), and incoherent layers. Generally, a layer should be treated as incoherent only if it is much much larger than the light wavelength, and if you have strong reason to believe that Fabry--P\'{e}rot fringes within that layer are getting averaged out.

As soon as light enters an incoherent layer, its phase information is thrown out, and only its intensity is remembered. This approach does not allow partial coherence, it's all or nothing! If that's not good enough, you can always fall back on the universally-valid method of running coherent simulations and averaging the results, as mentioned above.\footnote{There are more sophisticated methods for dealing with incoherence than the simple one used here; see Refs.~\cite{harbecke_coherent_1986, katsidis_general_2002} for example.}

\subsection{Calculation method}

\begin{figure}[htb]
\centering
\includegraphics[width=0.5\textwidth]{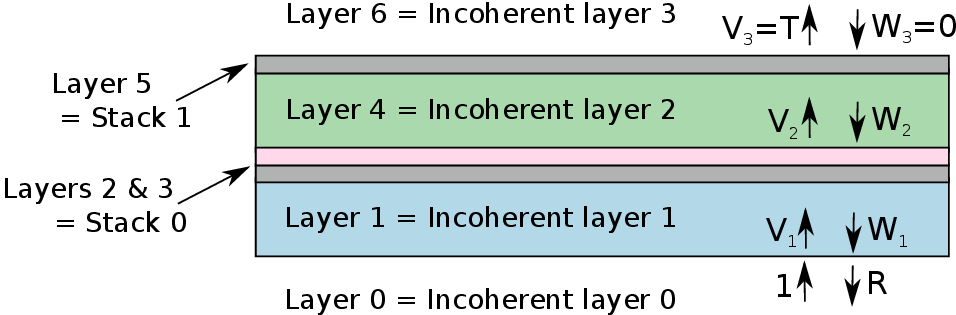}
\caption{Variable definitions related to the incoherent calculation program. A ``stack'' is one or more consecutive coherent layers. Note the three numbering systems: Each layer has a layer index, each incoherent layer has an incoherent layer index, and each stack has a stack index. $V_i,W_i$ are power flows (note the capital letters, not to be confused with the amplitudes $v_i,w_i$ in Fig.~\ref{vwdefn}).\label{incoherentfig}}
\end{figure}

We have a number of incoherent layers 0,1,...,$N-1$. Let $V_i$ be the forward propagation \emph{power} and $W_i$ be the backwards \emph{power} at the \emph{beginning} of the $i$th incoherent layer. (Capital letters to distinguish from $v,w$, the amplitudes in the coherent algorithm, see previous section.) Let $X_i$ and $Y_i$ be forward and backwards power at the \emph{end} of the $i$th incoherent layer. ($X_i$ and $Y_i$ are not explicitly calculated in the tmm software.) Let $T_{i,j}$ be transmissivity from the $i$th to $j$th incoherent layer (where $j=i\pm 1$, and $R_{i,j}$ the reflectivity. Then:
$$Y_i = X_i R_{i,i+1} + W_{i+1} T_{i+1,i}$$
$$V_{i+1} = X_i T_{i,i+1} + W_{i+1} R_{i+1,i}$$
\beq \smat X_i \\ Y_i \emat = \frac{1}{T_{i,i+1}} \smat 1 & -R_{i+1,i} \\  R_{i,i+1} & T_{i+1,i}T_{i,i+1} -  R_{i+1,i}R_{i,i+1} \emat \smat V_{i+1} \\ W_{i+1} \emat.\ee
Let $P_i$ be the fraction of light that passes successfully through layer $i$ (in a single pass) without getting absorbed, calculated by
\beq P_i = e^{-\alpha d_i}, \quad \alpha = \frac{4\pi \Im[n_i \cos \theta_i]}{\lambda_{vac}}\ee
where $d_i$ is the layer thickness. Then:
\beq \smat V_i \\ W_i \emat = \smat 1/P_i & 0 \\ 0 & P_i \emat \smat X_i \\ Y_i \emat\ee
Define the matrices $L_n$ by
\beq L_n =  \frac{1}{T_{i,i+1}} \smat 1/P_i & 0 \\ 0 & P_i \emat \smat 1 & -R_{i+1,i} \\  R_{i,i+1} & T_{i+1,i}T_{i,i+1} -  R_{i+1,i}R_{i,i+1} \emat\ee
for $n=1,\ldots,N-1$. Now we want the matrix relating the powers entering the structure to the powers exiting, i.e.:
\beq \smat 1 \\ R \emat = \tilde{L} \smat T \\ 0 \emat.\ee
Then the formula for $\tilde{L}$ is
$$\tilde{L} = \frac{1}{T_{0,1}} \smat 1 & -R_{1,0} \\  R_{0,1} & T_{1,0}T_{0,1} -  R_{1,0}R_{0,1} \emat L_1 L_2 \cdots L_{N-1} = \smat \tilde{L}_{00} & \tilde{L}_{01} \\ \tilde{L}_{10} & \tilde{L}_{11} \emat$$
\beq T = 1/\tilde{L}_{00}, \quad R = \tilde{L}_{10}/\tilde{L}_{00}\ee

\subsection{Absorption profile, Coherence length}

Absorption as a function of depth for incoherent layers is not implemented in the tmm software; this section explains why.

Calculating the absorption profile within an ``incoherent'' layer is not simple to do correctly. If you look up close, the absorption as a function of position would be oscillatory near an interface due to interference between the incoming and outgoing beams; with the oscillations gradually dying down into a smooth exponential farther away from the interface. The ``coherence length'' describes how far from the interface you need to go before the oscillations die down. For example, if the incoherence is caused by using a not-quite-monochromatic light source, the coherence length would be related to the bandwidth of the light.

If you are only interested in calculating the \emph{total} amount of light absorbed in each layer, it turns out that you do not need to know the coherence length!! More precisely, the coherence length does not affect the total absorption in (and transmission through) an incoherent layer under two assumptions (which are usually satisfied): (1) The coherence length is large compared to a wavelength; (2) The coherence length is small compared to the layer thickness.

This is an example of the more general mathematical fact that when you have sinusoidal oscillations that gradually die away, their integral is independent of the precise decay properties. Here is an example: $\int_0^\infty e^{ikx} e^{-\alpha x} dx = \frac{1}{-ik+\alpha} \approx \frac{1}{-ik}$; the integral is approximately independent of $\alpha$ as long as $\alpha \ll k$, i.e. as long as the decay length is much larger than the oscillation length.

That's the reason that you are not prompted to input coherence lengths in any of the calculations above.

On the other hand, if you want to calculate absorption as a function of depth in an incoherent layer, you \emph{do} need to know exactly what the coherence length is.

It is generally hard to know a coherence length \emph{quantitatively}. Therefore absorption as a function of depth is not implemented for incoherent layers in the tmm software. If you want to see absorption as a function of depth for an incoherent layer, you need to use the coherent program and average over slightly varying wavelengths / thicknesses / angles (as appropriate).

\section{Acknowledgments}

I thank Francis Loignon-Houle, John Honig, Fernando Stefani, Noah Rubin, Yinsheng Guo, Omer Luria, Fatemeh Edalatfar, Akira Okumura, and especially Mikhail Kats for helpful feedback and corrections.

\bibliographystyle{unsrt}
\bibliography{biblio}

\begin{thebibliography}{1}

\bibitem{harbecke_coherent_1986}
B.~Harbecke.
\newblock {Coherent and incoherent reflection and transmission of multilayer
  structures}.
\newblock {\em Applied Physics B}, 39(3):165--170, 1986.

\bibitem{katsidis_general_2002}
Charalambos~C. Katsidis and Dimitrios~I. Siapkas.
\newblock {General transfer-matrix method for optical multilayer systems with
  coherent, partially coherent, and incoherent interference}.
\newblock {\em Applied Optics}, 41(19):3978, 2002.

\bibitem{nistad_causality_2008}
Bertil Nistad and Johannes Skaar.
\newblock {Causality and electromagnetic properties of active media}.
\newblock {\em Physical Review E}, 78(3):036603, 2008.

\bibitem{chen_negative_2005}
Yi-Fan Chen, Peer Fischer, and Frank~W. Wise.
\newblock {Negative refraction at optical frequencies in nonmagnetic
  two-component molecular media}.
\newblock {\em Physical Review Letters}, 95(6):067402, 2005.

\end{thebibliography}

\newpage

\appendix

\section{Appendix: Sign convention for reflection amplitude \label{appendixrp}}

A common point of confusion for students is the sign convention for reflection amplitude in the Fresnel equations. For p polarization in particular, half of textbooks use one sign convention, the other half use the opposite one.\footnote{For example, the following textbooks use ``Convention A'' (as defined shortly): Jackson (Eq. (7.41)), Hecht (Eq. (4.38)), Zangwill (Eq. (17.34)); whereas the following textbooks use ``Convention B'': Feynman Lectures on Physics (Eq. 33.8), Griffiths (3rd edition, Eq. (9.109)), Lipson-Lipson-Lipson (Eq (5.42)).} The confusion more specifically is that it does not seem like it should be a convention (i.e., arbitrary choice) at all! After all, light interferes with its reflection. So the relative phase between light and its reflection does not seem like it should be arbitrary; it should have a right and wrong answer.

We start by asking: why is there a sign convention in the first place? The electric field vector of a wave is unambiguous, but if we want to write that vector as a scalar amplitude times a unit vector, then there are two ways to do that, because we can flip the sign of both the amplitude and the unit vector. So two ways of formulating the equations are (cf. Eq.~(\ref{Eq:PWaves})):
\begin{samepage}
\begin{center} {\bf \underline{Convention A (p-polarization)}} $\qquad$ {\bf \underline{Convention B (p-polarization)}} \end{center}
\vspace{-4mm}
\begin{align*}
\E_f = E_f(-\sin \theta \z +  \cos \theta \x) \quad\qquad &\qquad\quad \E_f = E_f(-\sin \theta \z +  \cos \theta \x) \\
\E_b = E_b(- \sin \theta \z -  \cos \theta \x) \quad\qquad &\qquad\quad \E_b = E_b( \sin \theta \z +  \cos \theta \x)
\end{align*} \end{samepage}
Again, the vectors $\E_f$ and $\E_b$ are the same for everyone, but different conventions will give different signs for $E_b$, and hence different signs for scalar quantities like the reflection amplitude $E_b/E_f$.

For a concrete example, consider light in air reflecting off glass at normal incidence. The electric field switches sign (changes phase by $\pi$), while the magnetic field keeps the same sign (changes phase by 0). Related to this, there are two ways to define the relative phase of oppositely-propagating light beams, the one based on whether the magnetic field is in phase or not (``Convention A''), and the one based on whether the electric field is in phase or not (``Convention B''). These two conventions line up with the two possible signs of $r_p$. In Convention A, $r_p > 0$ for light in air reflecting off glass at normal incidence; in Convention B, $r_p < 0$ in the same situation.

At normal incidence, there is a compelling reason to prefer Convention B: This is the convention based on whether or not the electric field of the incident and reflected light are in phase. The electric field is generally more important for light-matter interaction than the magnetic field. Relatedly, this gives $r_s = r_p$ at normal incidence, which neatly agrees with the fact that s and p are equivalent at normal incidence. (Everyone uses the same sign convention for $r_s$.)

However, at glancing angle, there is an equally compelling reason to prefer Convention A! At glancing angle, light and its reflection are traveling in the same direction, so there \emph{is} a unique and natural way to say whether the waves are in or out of phase at the interface. For example, if $r=-1$ at glancing angle, then that intuitively suggests that the reflected light is equal and opposite the incident light, so we expect destructive interference at the interface (both zero electric field and zero magnetic field). Convention A agrees with that expectation.

Again, there is no right or wrong convention, but we still have to pick one. So in this document we use Convention A.

\section{Appendix: \texorpdfstring{$R$ and $T$}{R and T} with an absorptive starting medium \label{appendixR}}

In this section, we discuss in more detail how we are defining $R$ and $T$, and why this can lead to unexpected results like $T>1$ when the starting medium is absorptive.

\begin{figure}[htb]
\centering
\includegraphics[width=0.5\textwidth]{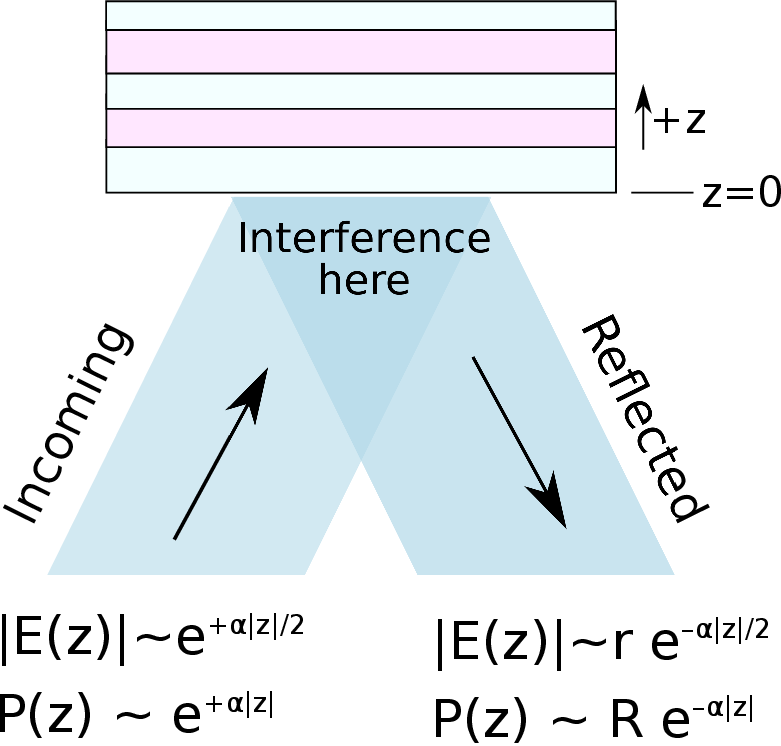}
\caption{How the reflected power $R$ is defined. Electric field $E$ and power transported $P$ are shown.\label{Rfig}}
\end{figure}

As shown in Fig.~\ref{Rfig}, I define $R$ and $T$ by the following operation:
\begin{itemize}
\item Assume that the light starts out a large distance $L$ behind the first interface, with power $P_1$ flowing towards the interface.
\item The light travels the distance $L$, bounces off, then travels the same big distance $L$ back through the initial medium, at which point it now has power $P_2$...
\item $R$ is defined by $R=e^{2\alpha L} P_2 / P_1$, in the limit $L\rightarrow \infty$. (The exponential factor here cancels out the absorption in the initial medium, so that this limit exists.) (We don't really need $L \rightarrow \infty$; all that really matters is that $L$ is big enough that it is beyond the area where the incoming and outgoing beams have coherent interference.)
\item Similarly, let $P_3$ be the flowing away from the last interface (immediately after the interface)...
\item $T$ is defined as $T =e^{\alpha L} P_3 / P_1$, in the limit $L \rightarrow \infty$.
\end{itemize}

(Are these good ways to define $R$ and $T$? Well, it depends on what you're trying to do. For example, if you are putting an antireflective coating on tinted glass, these are great definitions. They make it very easy to calculate the overall reflection and transmission. My incoherent tmm calculation is based on situations like that. So, I like these definitions, although I admit that they are not the only possible definitions.)

We obviously expect $R=|r|^2$ here (as usual), and that's correct with this definition.

The interesting thing---which had me confused at first---is that the normalized Poynting vector passing through the first interface is \emph{not} necessarily equal to $(1-R)$, as one would expect (energy 1 moving forwards minus energy $R$ moving backwards). Likewise, it is possible to have $R+T\neq 1$ for an interface between two semi-infinite media. This strange situation only comes up when the starting semi-infinite medium is absorbing. Why does this happen?

To get the exact formulas for Poynting vector at the initial interface, called \verb=power_entering= in the program, we just plug into the normal Poynting vector formula with $E_f = 1$ and $E_b = r$, to get:
\begin{center}\underline{s-polarization:}\end{center}
$$\text{power entering} = \frac{\Re \left[ n_0 \cos \theta_0  (1 + r^*) (1 -r) \right] }{\Re \left[ n_0 \cos \theta_0 \right] } = (1-R) + 2 \Im[r]\frac{\Im[n_0 \cos \theta_0]}{\Re[n_0 \cos\theta_0]}$$

\bskip\begin{center}\underline{p-polarization:}\end{center}
$$\text{power entering} = \frac{\Re \left[ n_0 \cos \theta_0^* (1 +r) (1 - r^*) \right] }{\Re \left[ n_0 \cos \theta_0^* \right] } = (1-R) - 2 \Im[r]\frac{\Im[n_0 \cos \theta_0^*]}{\Re[n_0 \cos\theta_0^*]}$$

The first term is what we expect, the second term is strange. We naively expect the Poynting vector for $z<0$ in Fig.~\ref{Rfig} to satisfy
\be{Snoninterfered} \S(z)\cdot \z = e^{\alpha |z|} - R e^{-\alpha |z|} \qquad \text{[formula for without wave  interference]} \quad \ee
where the first term comes from the incoming wave and the second term from the reflected wave. It is fine to demand this when the beams do not overlap, but we \emph{CANNOT} use this expression in the ``Interference here'' triangle region of Fig.~\ref{Rfig}. Instead, there is interference between the forward- and backward-moving waves, which causes oscillations in the absorption profile (``hot-spots'' and ``nodes'') (see Fig.~\ref{spatiallyresolvedexample}), so there are corresponding oscillations in the Poynting vector (it's a bit hard to see them in Fig.~\ref{spatiallyresolvedexample} but they're there).  A bit of extra energy is flowing from the nodes to the nearby hot-spots. Thanks to these oscillations, the Poynting vector right at the edge before the start of the structure may not be $1-R$.

If $r$ is real, then there is a node or hot-spot of absorption right at the interface. It turns out that this sort of corresponds to having an integer number of oscillatory cycles, so the oscillations do not affect the power passing through the interface. $R+T=1$ is still valid. But if $r$ is complex, then you have an extra bit of absorption, or deficit of absorption, compared to the non-oscillating baseline expectation of Eq.~(\ref{Snoninterfered}).

Instead of $R+T=1$, the formula is:
$$R+T\pm \text{(extra bit or deficit of absorption from how the oscillations cut off)} = 1$$
Again, this comes from the fact that $R$ is defined by Eq.~(\ref{Snoninterfered}), which is not valid when there is interference.

To verify that $R=|r|^2$ is the correct expression to use, we use the Poynting vector formula, plugged in at an arbitrary depth $z<0$, using $E_f = \exp(2\pi i n z \cos \theta / \lambda_{vac})$ and $E_b = r \exp(-2\pi i n z \cos\theta/\lambda_{vac})$. I'll just do the s-polarized case:
\bea
\S(z)\cdot\z
&=& \frac{\Re[n_0 \cos \theta_0 (E_f^* + E_b^*)(E_f-E_b)]}{\Re[n_0 \cos \theta_0]}\\
&=& \left(e^{-4\pi z \Im[n_0 \cos \theta_0]/\lambda_{vac}}-|r|^2 e^{+4\pi z \Im[n_0 \cos \theta_0]/\lambda_{vac})}\right) \; +\\
&\;& + \left( \; 2\frac{\Im[n_0 \cos\theta_0]}{\Re[n_0 \cos \theta_0]} \Im [r e^{4\pi i z\Re[n_0  \cos \theta_0]/\lambda_{vac}}] \right)\\
\eea
The first term corresponds exactly to Eq.~(\ref{Snoninterfered}) with $R=|r|^2$, and the second term is a sinusoidal oscillation corresponding to interference. When the beams stop overlapping (i.e., below the triangle in Fig.~\ref{Rfig}), the oscillation term goes away but the other term remains.

In the program, I use the variable ``\verb=power_entering='' to describe the net power entering the structure, i.e. the Poynting vector at the front of the first layer. For an interface between two semi-infinite media with light coming from just one side, \verb=power_entering= is always equal to $T$. When the incident semi-infinite medium has real refractive index, \verb=power_entering= is always equal to $1-R$.

\subsection{Accounting for this effect in the incoherent calculation}

One of the things I want to compute in the incoherent calculation is how much light gets absorbed in each layer. Part of that absorption is the ``extra'' absorption due to the oscillation cut-off at the interface.

At the interface between two incoherent layers, say 0 and 1, let's say the power flows on the two sides are $P_{f,0},P_{f,1},P_{b,0},P_{b,1}$, where f and b stand forward-moving and backward-moving. The important thing to remember is that the net power actually crossing the interface is \emph{exactly} $P_{f,0}T_{01} - P_{b,1}T_{10}$. Why? Because the transmitted light beams have no funny corrections due to oscillations; they have  nothing to \emph{coherently} interfere with them.

Therefore, the ``extra'' absorption near the interface, not accounted for in the exponential decay of the waves, is exactly equal to
$$(P_{f0}-P_{b0}) - (P_{f0}T_{01} - P_{b1}T_{10}) = P_{f0}(1-R_{01}-T_{01})$$
extra near-interface absorption on the 0 side, and likewise
$$P_{b1}(1-R_{10}-T_{10})$$
extra near-interface absorption on the 1 side.

\section{Appendix: Stimulated emission \label{appendixgain}}

With stimulated-emission (a.k.a. ``gain'' or ``active'') media, there is a possibility for confusion. It turns out that Maxwell's equations always have a unique \emph{finite} steady-state solution, and the algorithm described herein will always find this solution. But this solution may be unphysical! There is, after all, another possibility: The system may be unstable, with fields exponentially growing (until the gain saturates). So you can find various papers exploring surprising aspects of Fresnel reflection and refraction with gain---but where the results are all nonsense, because they are exploring unphysical solutions. A very helpful paper in this area is Ref.~\cite{nistad_causality_2008}.

The same paper also explains why knowing $n$ at one wavelength is not enough information to do a multilayer fresnel analysis when the initial or final layer has stimulated emission: The whole wavelength-vs-$n$ dispersion is required to figure out which choice of arcsine to use (cf. Section~\ref{branchcuts}).

If a medium has gain at $\lambda_1$ but loss at $\lambda_2$, it can require you to use the unexpected choice of arcsine even at $\lambda_2$! An example along those lines is constructed in Ref.~\cite{chen_negative_2005}, where a certain dielectric function at a certain wavelength has negative refraction, even though it has (at that wavelength) neither negative permittivity nor negative permeability nor gain.

\section{Appendix: Branch cuts \label{appendixbranchcuts}}

Snell's law gives $\theta_i = \arcsin(n_0 \sin(\theta_0)/n_i)$. However, the arcsine function is ambiguous--it has branch cuts. How to get the right $\theta$?? Remember here, $n_i$ may be an arbitrary complex number, ideally the program will work even for unusual cases like negative-index materials ($\Re n <0$) or stimulated-emission media ($\Im n<0$).

Actually, we never care about $\theta$ itself, just $\sin \theta$ and $\cos \theta$. The sine has no ambiguity:
$$\sin \theta_i = \frac{n_0\sin \theta_0}{n_i}$$
The cosine is more problematic, because there are two choices consistent with Snell's law:
$$\cos \theta_i = \pm \frac{1}{n_i} \sqrt{n_i^2 - (n_0 \sin \theta_0)^2}$$
Do we want the $+$ or $-$?? In other words, we can pick between two angles $\theta$ and $\pi-\theta$.

See Section~\ref{branchcuts} for an explanation of what the choice \emph{really} means, and more importantly, why it only matters in the starting and ending semi-infinite layers, but doesn't matter in the intermediate, finite-thickness layers.

\subsection{Computer implementation}

I do \emph{not} recommend using $\theta_i = \arcsin(n_0 \sin(\theta_0)/n_i)$ and hoping to wind up with the right $\theta$. That often works, but not always, at least not always in all programming languages. It depends on details of the arcsine branch cut implementation. In Python/SciPy, I found that this gives the wrong $\theta_i$ for total internal reflection when $\theta_0>0$. Even if it \emph{seems} to always work, you are vulnerable to things like rounding errors pushing you to the other side of the branch cut, or changes in the arcsine definition when you upgrade your software, or whatever---it's not a robust solution.

A much better idea is to calculate $\theta = \arcsin(n_0 \sin(\theta_0)/n_i)$, then \emph{check} that this is the right angle (using the criteria below), and if not, use $\pi-\theta$ instead. (As mentioned above, you only need to check the starting and ending semi-infinite layers.)

\subsection{Case that \texorpdfstring{$\Im n>0, \quad \Re n > 0$}{Im n > 0, Re n > 0}}

For an absorbing material ($\Im n > 0$), a clear requirement is that $\Im(n \cos \theta)>0$. That way, $ \Im k_z > 0$, so the $E_f$ wave in the medium decays rather than amplifying. Another clear requirement is that $R>1$ or $T<0$ cannot occur. This translates to $\Re[n \cos\theta]\geq 0$ for s-polarization and $\Re[n \cos \theta^*]\geq 0$ for p-polarization. This amounts to the same thing as saying that the Poynting vector associated with an $E_f$ wave should point forward not backwards.

Important question: Are these two ``clear requirements'' consistent with each other. Yes!

{\bf Theorem:} If we choose the $\theta$ with $\Im(n \cos \theta)>0$, then it will also be true that $\Re[n \cos\theta] > 0$.

{\bf Proof:} As above,
$$n_i \cos \theta_i = \pm \sqrt{n_i^2 - (n_0 \sin \theta_0)^2}$$
Given that $\Im n_i>0$ and $\Re n_i > 0$, it follows that $\Im n_i^2 >0$. Since $n_0 \sin \theta_0$ is real (the wave intensity is assumed to be uniform in the lateral direction), $(n_i^2 - (n_0 \sin \theta_0)^2)$ also has a positive imaginary part. Therefore, its square root is in the first or third quadrant of the complex plane. That finishes the proof.

{\bf Theorem:} If we choose the $\theta$ with $\Im(n \cos \theta)>0$, then it will also be true that $\Re[n \cos\theta^*] > 0$.

{\bf Proof:}
$$\Re[n_i \cos \theta_i^*] = \Re[n_i^* \cos \theta_i ] = \pm \frac{n_i^*}{n_i}\sqrt{n_i^2 - (n_0 \sin \theta_0)^2}$$
Let $\phi$ with $0<\phi<\pi/2$ be the complex phase of $n_i$. We have
$$\arg \frac{n_i^*}{n_i} = -2\phi$$
Using the fact that $(n_0 \sin \theta_0)^2$ is a nonnegative real number,
$$\arg n_i^2 = 2\phi, \quad 2\phi \leq \arg (n_i^2 - (n_0 \sin \theta_0)^2) < \pi$$
If we choose the square-root with positive imaginary part,
$$\phi \leq \arg \sqrt{n_i^2 - (n_0 \sin \theta_0)^2} < \pi/2$$
Therefore,
$$-\pi/2 < -\phi \leq \arg \left[ \frac{n_i^*}{n_i} \sqrt{n_i^2 - (n_0 \sin \theta_0)^2}\right] < -2\phi+\pi/2 < \pi/2$$
That finishes the proof.

\subsection{Case that \texorpdfstring{$\Im n=0, \quad \Re n > 0$}{Im n = 0, Re n > 0}}
As before,
$$\cos \theta_i = \pm \frac{1}{n_i} \sqrt{n_i^2 - (n_0 \sin \theta_0)^2}$$
There are three cases: Total internal reflection where $n_0\sin\theta_0>n_i$ and $\cos \theta_i$ is pure imaginary; the ordinary case where $n_0 \sin \theta_0 < n_i$ and $\cos \theta_i$ is pure real, and the boundary case where $n_0 \sin \theta_0 = n_i$ and $\cos \theta_i = 0$. The third one has no ambiguity because $\cos \theta_i = -\cos \theta_i$. Let's look at the other two cases.

\subsubsection{Case \texorpdfstring{$\Im n=0, \quad \Re n > 0$}{Im n = 0, Re n > 0}, Total internal reflection}
Here, $\Re[n_i \cos\theta_i] = 0$ and $\Re [n_i^* \cos \theta_i^*]=0$, so no need to worry about the sign of the Poynting vector or $R>1$ or $T<0$. The only requirement is that the wave decay rather than amplify, i.e.
$$\Im(n_i \cos \theta_i)>0$$

\subsubsection{Case \texorpdfstring{$\Im n=0, \quad \Re n > 0$}{Im n = 0, Re n > 0}, Normal refraction}

Here, $\Im(n_i \cos \theta_i)=0$, so we get no information from whether the wave is decaying or amplifying. The only criterion is $\Re[n_i \cos\theta_i] > 0$ and $\Re [n_i^* \cos \theta_i^*]>0$ (meaning the Poynting vector points forward, $R<1$, $T>0$). In this case it simplifies to $n_i \cos\theta_i > 0$.



\subsection{Case that \texorpdfstring{$\Im n<0, \quad \Re n < 0$}{Im n < 0, Re n < 0}}

This is \emph{not} stimulated emission, despite $\Im n<0$. It is absorption. [Remember, with $\Re n<0$, the direction the wave is ``really moving'' (the direction of the Poynting vector) is opposite the direction of the wavevector. When $\Im n<0$, the wave is amplifying in the direction of the wavevector, so it's ``really'' decaying.]

Therefore the required criteria are the same as for ordinary absorbing media with $\Im n>0$ and $\Re n > 0$: $\Im(n \cos \theta)>0$ (the $E_f$ wave decays rather than amplifies),  $\Re[n \cos\theta]\geq 0$ for s-polarization and $\Re[n \cos \theta^*]\geq 0$ for p-polarization (the $E_f$ wave carries energy forwards and $R<1$ and $T>0$.)

If we flip the sign of $n_i$, we do not affect $\sqrt{n_i^2 - (n_0 \sin \theta_0)^2}$, so we do not affect $\cos \theta_i$ (up to a possible sign-flip) nor do we affect $(n_i \cos \theta_i)$ (up to a possible sign-flip). Therefore the proof is exactly the same as before that the two requirements are consistent with each other.

\subsection{Everything else}

As mentioned above, active media cannot be analyzed in this way, because knowing $n$ at one wavelength is not enough information to determine which solution is which---see Ref.~\cite{nistad_causality_2008}.

There are other cases too, like $n=0$, which I have not looked into.







\end{document}